\documentclass[pra,twocolumn,a4paper,showpacs,floatfix,superscriptaddress]{revtex4}  

\usepackage{graphicx}
\usepackage{amsmath}
\usepackage{amssymb}
\usepackage{amsthm}
\usepackage{color}

\graphicspath{{./IMAGES/}}

\begin{document}
\title{Localization of massless Dirac particles via spatial modulations of the Fermi velocity}

\author{C. A. Downing}
\email[]{downing@ipcms.unistra.fr}
\affiliation{Universit\'{e} de Strasbourg, CNRS, Institut de Physique et Chimie des Mat\'{e}riaux de Strasbourg, UMR 7504, F-67000 Strasbourg, France}

\author{M. E. Portnoi}
\email[]{m.e.portnoi@exeter.ac.uk}
\affiliation{School of Physics,
University of Exeter, Stocker Road, Exeter EX4 4QL, United Kingdom}
\affiliation{International Institute of Physics, Universidade Federal do Rio Grande do Norte, Natal - RN, 59078-970, Brazil}

\date{\today}

\begin{abstract}
The electrons found in Dirac materials are notorious for being difficult to manipulate due to the Klein phenomenon and absence of backscattering. Here we investigate how spatial modulations of the Fermi velocity in two-dimensional Dirac materials can give rise to localization effects, with either full (zero-dimensional) confinement or partial (one-dimensional) confinement possible depending on the geometry of the velocity modulation. We present several exactly solvable models illustrating the nature of the bound states which arise, revealing how the gradient of the Fermi velocity is crucial for determining fundamental properties of the bound states such as the zero-point energy. We discuss the implications for guiding electronic waves in few-mode waveguides formed by Fermi velocity modulation.
\end{abstract}

\maketitle

\section{\label{intro}Introduction}

It is a truth universally acknowledged, that a single electron in possession of a gapless, linear spectrum, must be in want of a bound state \cite{Klein}. There is a considerable resurgence in the importance of such massless fermions in condensed matter physics due to the rise of so-called Dirac materials \cite{Wehling}, whose charge carriers behave according to quasi-relativistic wave equations. Celebrated examples in two dimensions include graphene or the surface states of topological insulators. An important property that is inherent to such Dirac particles is the absence of backscattering \cite{Katsnelson}, which whilst leading to large electron mobilities, presents a considerable difficulty in localizing Dirac electrons \cite{Trauzettel, Martino, Bardarson, Hartmann2010, Rozhkov, Giavaras2012, Zalipaev2013, Optimal, Mag2} and hence building practical digital devices with a well-defined on/off logical state \cite{Yung2013}.

One interesting method proposed to manipulate these somewhat elusive quasi-relativistic charge carriers is to consider systems with a spatially-varying Fermi velocity, $v_F = v_F(\mathbf{r})$ \cite{deJuan, Peres, Amorim}, such that so-called velocity barriers may form. The resulting ballistic electron transport in such systems has already been extensively studied \cite{Concha, Raoux, Krstajic, Pellegrino, Liu2013, Cheraghchi}, as has the effects of applying external electric \cite{Liu1} and magnetic fields \cite{Yuan2} and a superlattice geometry \cite{Esmailpour, Lima1, Lima2, Bezerra}. There are immediately apparent strong analogies in both acoustics and especially optics, where phenomena such as super-collimation has been envisaged \cite{Yuan, Wang1}. 

Energy-dependent Fermi velocity renormalization has already been seen in experiments with graphene at energies close to the Dirac point \cite{Elias, Siegel}. Here we instead consider the problem of Dirac particles that can be described with a spatially modulated Fermi velocity. This situation arises theoretically from both elasticity theory with tight-binding calculations, as well as quantum field theory in curved space \cite{Sturla}. Experimentally, a spatially dependent Fermi velocity may occur due to ripples in the material \cite{Martin, Polini}, the use of different substrates \cite{Luican, Hwang2012}, superlattices \cite{Park, Gibertini, Yan}, atomic scale defects induced by ion irradiation \cite{Tapaszto}, straining the material \cite{Huang2010, Jang}, by placing a grounded plane of metal nearby \cite{Raoux} or by judiciously applying a uniform electric field \cite{Diaz2017}. Indeed, a spatial dependence of the Fermi velocity has recently been observed in two different experiments \cite{Yan, Jang}. Moreover, our work is relevant for a large range of artificial Dirac materials, which advantageously allow one precise control over the velocity of the hosted Dirac-like particles. Examples of artificial Dirac systems include: cold atoms in an optical lattice \cite{Zhu2007}; flexural waves in thin plates \cite{Torrent2013}; microwaves in a lattice of dielectric resonators \cite{Bellec2013}; and plasmons in metallic nanoparticles \cite{Downing2017}. 

Previously, most of the theoretical attention on this topic has been focused on the scattering of two-dimensional (2D) massless Dirac fermions on square velocity barriers, either single \cite{Yuan}, double \cite{Liu2} or multiple \cite{Esmailpour, Lima1, Lima2}. Here we address the bound state problem for 2D Dirac particles and furthermore we consider non-square velocity distributions, which are arguably more realistic. Indeed, smooth electrostatic and magnetostatic potential barriers are known to lead to effects not found in their sharp barrier counterparts \cite{Stone2012, Mag}.

In this work, we reveal the dependence of the supported bound state energy levels on the velocity barrier parameters and find how the presence of a finite zero-point (ground state) energy is critically dependent on the gradient of the velocity barrier. For one-dimensional (1D) confinement, whilst exponentially growing velocity barriers have a finite threshold energy for the ground state, algebraically growing velocity barriers (growing at most linearly) have a vanishing ground state energy. This 1D geometry of velocity barrier acts as an electron waveguide for massless quasirelativistic particles \cite{Beenakker2009, Zhang2009, Myoung2011, Stone2012, Hartmann2014}, with the bound modes propagating along the channel created by the confining barrier. Such systems are of great interest to experimentalists hoping to realize electron optics based on Dirac materials such as graphene \cite{Chen2016}. For zero-dimensional localization (0D), we show how true confinement in radial velocity barriers acting as nanoscale quantum dots is indeed possible, and demonstrate how both an infinite velocity barrier and a smooth, algebraically growing radial barrier possess a nonvanishing zero-point energy. The proposed bound states may be revealed in experiments via quantum transport measurements, where the signature of a confined mode is a jump in the conductance.

It has been shown by Peres \cite{Peres} that in order to maintain Hermitian operators, the relevant (and Sturm-Liouville) 2D Dirac Hamiltonian for this problem is
\begin{equation}
\label{intro00}
 \hat{H} = \sqrt{v_F(\mathbf{r})} \pmb{\sigma} \cdot \mathbf{\hat{p}} \sqrt{v_F(\mathbf{r})},
\end{equation}
where $\pmb{\sigma} = (\sigma_x, \sigma_y)$ are the spin matrices of Pauli. Eq.~\eqref{intro00} acts on a two-component spinor wavefunction $\Phi (\mathbf{r})$, and the eigenvalues $E$ are found via $\hat{H} \Phi = E \Phi$. Continuity of the probability current leads to the following boundary condition at an interface $\mathbf{r} = \mathbf{R}$
\begin{equation}
\label{intro01}
 \sqrt{v_F(\mathbf{r})} \Phi (\mathbf{r}) \Bigr|_{\mathbf{r} = \mathbf{R}+\pmb{\delta}} = 
		\sqrt{v_F(\mathbf{r})} \Phi (\mathbf{r}) \Bigr|_{\mathbf{r} = \mathbf{R}-\pmb{\delta}},
\end{equation}
which is directly analogous to what occurs in heterostructures defined by a position-dependent mass \cite{Cole}. In what follows, we make the following assignment of the auxiliary spinor for convenience $\Psi (\mathbf{r}) = \sqrt{v_F(\mathbf{r})} \Phi (\mathbf{r})$. We solve Eq.~\eqref{intro00} for various spatial profiles of the Fermi velocity $v_F(\mathbf{r})$ to unveil the general properties of bound states in such systems, with the ultimate aim of proposing on/off logical states in Dirac materials.

The rest of this work is organized as follows. We study in Sec.~\ref{sec2} a series of different 1D velocity barriers giving rise to bound states. Localization in radially symmetric 2D velocity barriers is discussed in Sec.~\ref{sec3}. Finally, we draw some conclusions in Sec.~\ref{conc}.

\section{\label{sec2}Velocity barrier channels}

\begin{figure}[tb]
 \includegraphics[width=0.4\textwidth]{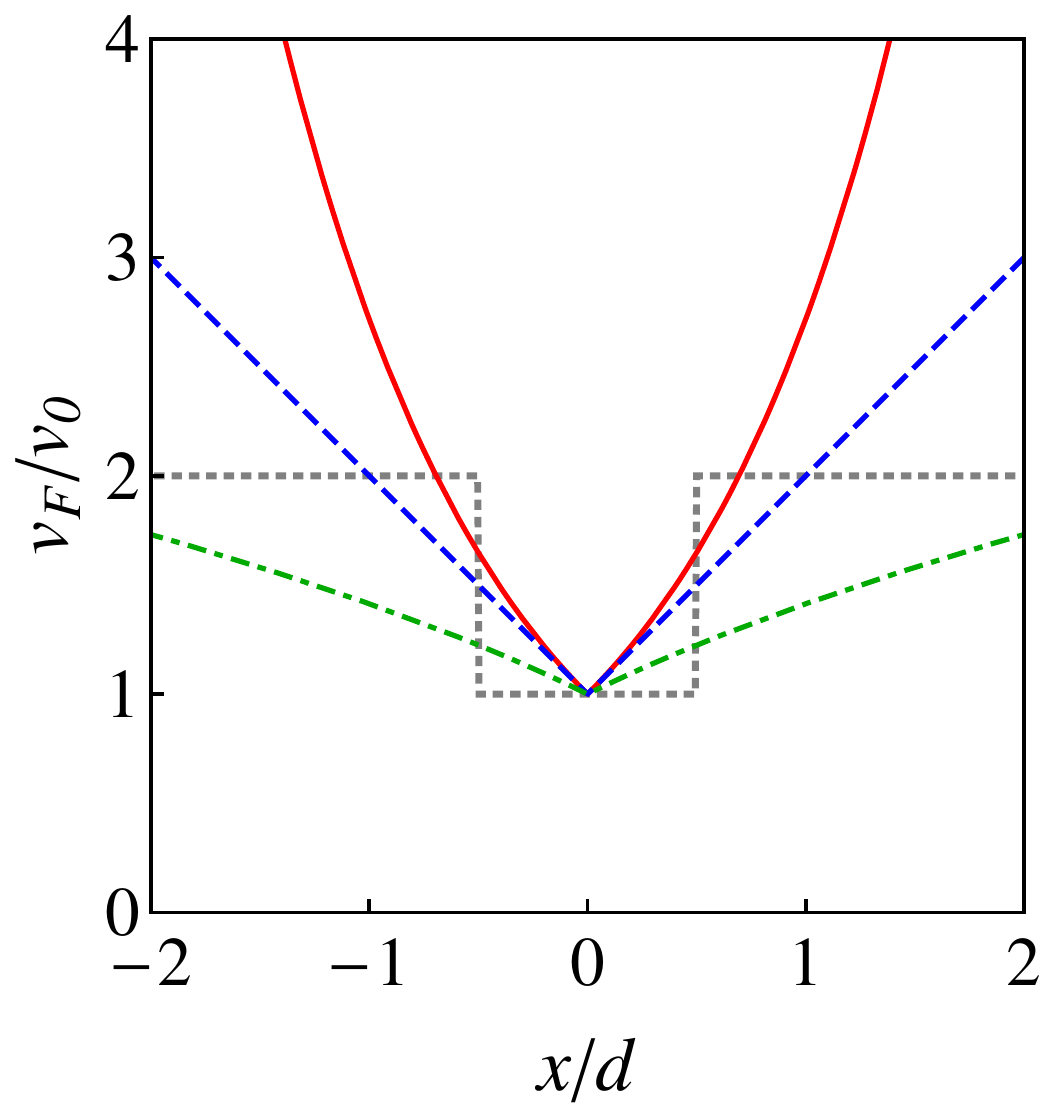}
 \caption{(Color online) Profiles of the considered 1D velocity barriers: the square (short-dashed gray line), the exponential (solid red line), the linear (long-dashed blue line) and the square root (dot-dashed green line) channels respectively. Here we take $v_1 / v_0 = 2$ for the square velocity barrier.}
 \label{fig1}
\end{figure}

We shall consider several toy models of 1D velocity barriers $v_F = v_F(x)$, each with drastically different spatial profiles, yet all unified by their integrability. Namely, we investigate the following 1D velocity  barrier channels: square (Sec.~\ref{subssec1}), exponential (Sec.~\ref{subssec2}), linear (Sec.~\ref{subssec3}) and square root (Sec.~\ref{subssec4}), as shown graphically in Fig.~\ref{fig1}. This range of models allows us to identify how the shape of the velocity barrier influences the properties of the bound states.

Working in Cartesian coordinates $(x, y)$, we begin by making the following ansatz for the auxillary spinor $\Psi (x,y) = (L_y)^{-1/2} e^{i q_y y} \left[ \psi_1(x), \psi_2(x) \right]^T$, due to translational invariance in the $y$ direction. Here $q_y$ is the wavenumber along the formed trench and $(L_y)^{-1/2}$ is the length of the material in the $y$-direction.

\subsection{\label{subssec1}The square velocity barrier}

\begin{figure}[tb]
 \includegraphics[width=0.4\textwidth]{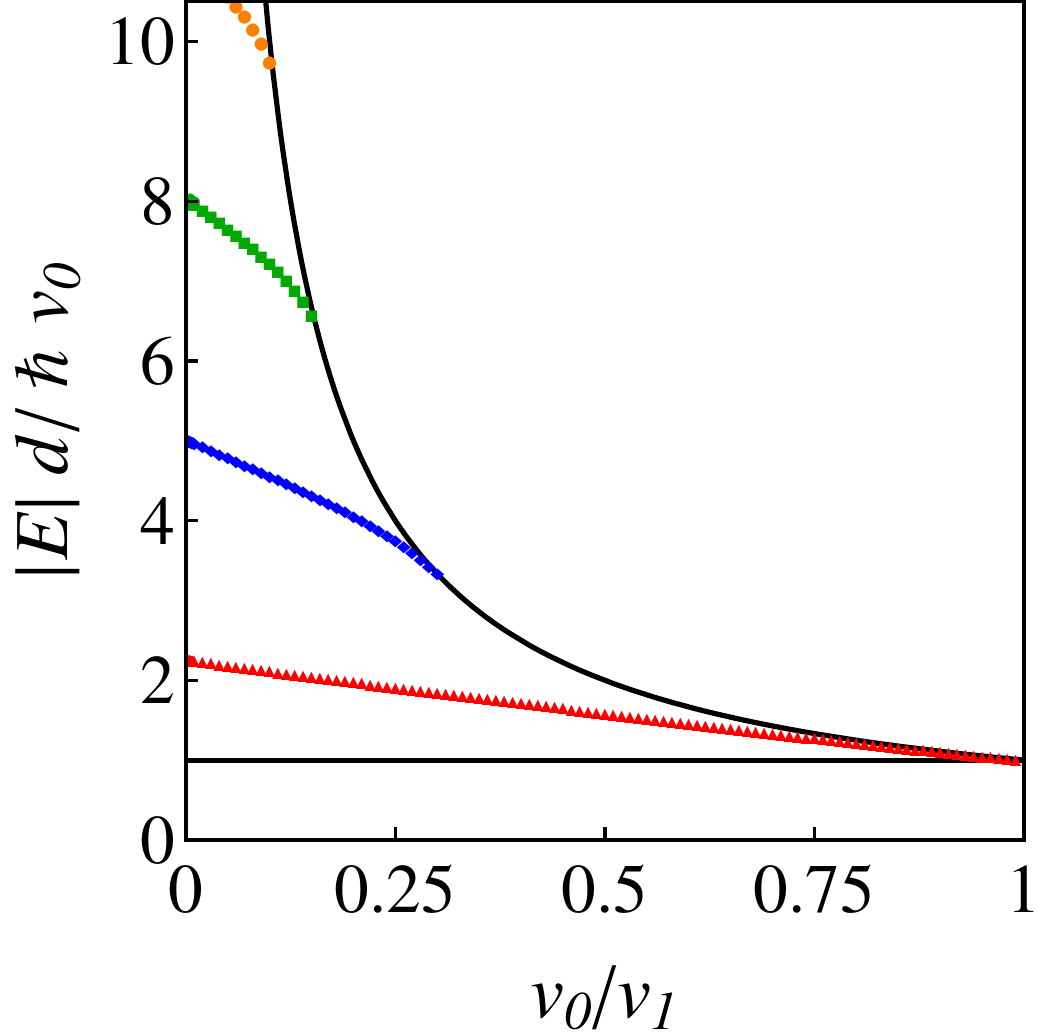}
 \caption{(Color online) Eigenenergies of massless Dirac fermions in a square velocity barrier, as a function of the barrier parameter $v_0 / v_1$, calculated via Eq.~\eqref{eqbox02}. The lowest four states can be seen, from the ground state (red triangular markers), to higher states (blue diamond markers, green square markers and orange circular markers respectively). The region of allowed bound states, arising from Eq.~\eqref{eqbox013343443}, is denoted by solid black lines. Here we take $q_y d = 1$.}
 \label{fig2}
\end{figure}

Firstly, we revisit the square model velocity barrier \cite{Peres, Concha, Raoux, Krstajic}. Whilst there has been much focus on scattering on such a barrier, instead we shall focus our investigation on the associated bound states that may be supported. We consider a square barrier of width $d$, given by
\begin{equation}
\label{eqbox00}
 v_F(x)= \begin{cases} v_0, \quad -\tfrac{d}{2} \le x \le \tfrac{d}{2},  \\
                     v_1, \quad x > \tfrac{d}{2}, \quad x < -\tfrac{d}{2},  \end{cases}
\end{equation}
where the velocity parameters satisfy $v_1 > v_0$, as plotted in Fig.~\ref{fig1} as the short-dashed gray line. Naturally appearing in this problem are the following wavenumbers
\begin{equation}
\label{eqbox01}
 \kappa = \left(q_y^2 - \left( \tfrac{E}{\hbar v_1} \right)^2 \right)^{1/2}, \quad k = \left(\left( \tfrac{E}{\hbar v_0} \right)^2 - q_y^2 \right)^{1/2},
\end{equation}
which arise when describing the spinor wavefunction in its evanescent and propagating stages respectively. These wavevectors automatically restrict the region of bound states to the fan
\begin{equation}
\label{eqbox013343443}
 \hbar v_0 |q_y| < |E| < \hbar v_1 |q_y|,
\end{equation}
since it is required that both $\kappa$ and $k$ are real quantities. The energies of the bound states in the square barrier follow from the Hamiltonian~\eqref{intro00} with the boundary condition Eq.~\eqref{intro01}, and are determined by the transcendental equation
\begin{equation}
\label{eqbox02}
 \tan (kd) = \frac{2 \frac{v_1}{v_0}\frac{\kappa}{k}}{1 + \left( 1 - \frac{v_1}{v_0} \right)^2 \left( \frac{q_y}{k} \right)^2 - \left( \frac{v_1}{v_0} \frac{\kappa}{k} \right)^2 }.
\end{equation}
We plot in Fig.~\ref{fig2} the eigensolutions of Eq.~\eqref{eqbox02}, as a function of the barrier strength parameter $v_0/v_1$, showing the four lowest-lying bound states. As is common in 1D barrier problems, there is always at least one bound state, even for arbitrarily weak barriers with $v_0/v_1 \to 1$. Upon taking this limit, one can see from Eq.~\eqref{eqbox02} that the left hand side of the equation will grow monotonically between $0 \le \tan (kd) \le q_y d \left( v_1^2/v_0^2 - 1\right)^{1/2}$. This left hand side will always be intercepted as the right hand side of Eq.~\eqref{eqbox02} shrinks to zero as $|E| \to \hbar v_1 |q_y|$ and hence a bound state solution is guaranteed.

In the ultrastrong barrier limit $v_1 \to \infty$, Eq.~\eqref{eqbox02} must be replaced with the equation $\tan (kd) + k/q_y = 0$. Now there are an increasing number of bound states $N$, which can be estimated from
\begin{equation}
\label{eqbox03}
 N = \left \lfloor \Upsilon \frac{q_y d}{\pi} \right \rfloor, \quad \Upsilon = \left( v_1^2/v_0^2 - 1\right)^{1/2},
\end{equation}
where $\left \lfloor ... \right \rfloor$ is the floor function. Inverting Eq.~\eqref{eqbox03} tells us the threshold velocity barrier strengths above which new bound states appear, via the approximate relation
\begin{equation}
\label{eqbox033}
 v_0/v_1 \simeq \left( 1 + \left( \frac{\pi N}{q_y d} \right)^2 \right)^{-1/2}.
\end{equation}
This feature of a changing number of bound states with a modulation of $v_0/v_1$ is shown in Fig.~\ref{fig2}. This phenomenon of losing successive bound states into the continuum as $v_0/v_1$ increases towards unity is superficially reminiscent of the `fall-into-the-center' phenomenon in relativistic quantum mechanics \cite{Zeldovich}. The analogue of this so-called atomic collapse effect in Dirac material physics (in 1D) sees the lowest lying states diving into the continuum as the band-gap of the Dirac material is reduced \cite{Hartmann2011, Collapse}. A notable distinction here is that whilst conventionally the bound states with fewest nodes are successively lost, for the square velocity barrier the nodeless ground state is always present and instead the highest lying (highly nodal) states are successively lost with increasing $v_0/v_1$. For example, in Fig.~\ref{fig2} and with $q_y d = 1$, the third, second and first excited states (orange circular markers, green square markers and blue diamond markers respectively) disappear one-by-one as $v_0/v_1$ increases at $v_0/v_1 \simeq 0.11, 0.16, \text{and}~0.30$ in turn, whilst the ground state (red triangular markers) persists even as $v_0/v_1 \to 1$.

In what follows, in Secs.~\ref{subssec2}-~\ref{subssec4}, we move on to investigating non-square velocity barriers, with an emphasis on the bound modes propagating along the formed Dirac electron waveguides with wavenumber $q_y >0$.

\subsection{\label{subssec2}The exponential velocity barrier}

\begin{figure}[tb]
 \includegraphics[width=0.4\textwidth]{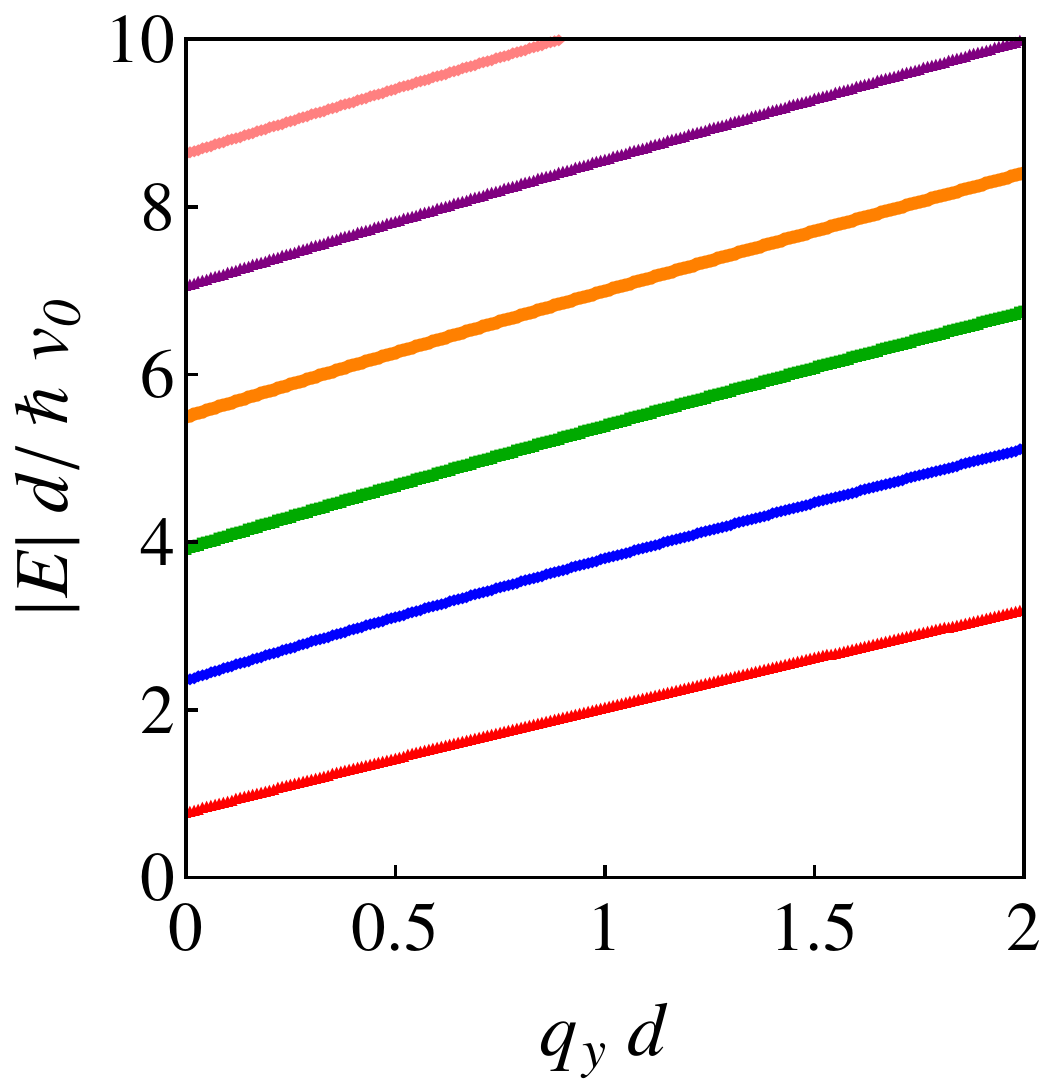}
 \caption{Progression of the six lowest bound states energies with traversal momentum $q_y d$, for massless Dirac particles in a cusp-like velocity barrier, calculated via Eq.~\eqref{eqcusp02}.}
 \label{fig3}
\end{figure}

As a first example of a smooth velocity barrier, let us study the exponential velocity barrier, plotted in Fig.~\ref{fig1} as the solid red line, and defined by
\begin{equation}
\label{eqcusp00}
 v_F(x)= v_0 e^{|x|/d}.
\end{equation}
Here $v_0$ is the minimal Fermi velocity, found at the center of the barrier, and $d$ is the length scale of the problem, from which arises the key dimensionless parameter
\begin{equation}
\label{eqcusp000}
 \lambda = \frac{|E|d}{\hbar v_0},
\end{equation}
which is useful for describing the eigenvalues. Upon solving the coupled equations formed from the Hamiltonian~\eqref{intro00}, one finds the following spinor wavefunction $\psi(x)$ in region I $(x>0)$ with the help of the independent variable change $\xi = \lambda e^{-x/d}$
\begin{equation}
\label{eqcusp01}
\psi_{I}(x) 
 = \tfrac{c_{I}}{\sqrt{d}} e^{-x/2d}
\left(
 \begin{array}{c}
 J_{q_y d + 1/2} \left( \xi \right) \\ \text{sgn} (E) J_{q_y d - 1/2} \left( \xi \right)
 \end{array}
\right),
\end{equation}
in terms of the Bessel function of the first kind $J_{\alpha}(\xi)$ and where $c_{I}$ is some constant. The solution in region II $(x<0)$ is found by interchanging the top and bottom wavefunction components, and making the replacements $x \to -x$ and $c_I \to c_{II} = \pm \text{sgn} (E) c_I$. Applying the boundary condition Eq.~\eqref{intro01} at the interface $x=0$, one finds the spectrum of bound states is determined via the transcendental equation 
\begin{equation}
\label{eqcusp02}
 J_{q_y d - 1/2} \left( \lambda \right) = \pm J_{q_y d + 1/2} \left( \lambda \right),
\end{equation}
where the $\pm$ corresponds with the $\pm$ in the definition of $c_{II}$. Eq.~\eqref{eqcusp02} can be solved with standard root-finding methods or indeed graphically, the result of which is shown in Fig.~\ref{fig3}. There we plot the six lowest lying bound states as a function of transversal wavevector $q_y d$, revealing an approximately linear dependence between energy level and transversal momentum, as well as an approximately constant level spacing.

One also notices there is a threshold magnitude of energy at which the first bound state appears. This zero-point energy can be quantified by taking the limit $q_y d \to 0$ in Eq.~\eqref{eqcusp02}, such that one arrives at the analytic expression 
\begin{equation}
\label{eqcusp03}
 \lambda_{n, \pm} = \pi \left( n \pm \tfrac{1}{4} \right),
\end{equation}
where $n$ is a nonnegative (positive) integer when the $+ (-)$ sign is taken. Explicitly, the ground state energy is $\lambda_0 = \pi / 4 \simeq 0.785$. In this small wavevector limit, the energy level separation is a universal constant $\Delta \lambda = \pi /2 \simeq 1.57$.

The characteristics of this velocity barrier suggest it can act as a few-mode, Dirac electronic waveguide in direct analogy to the channeling of photons along optical fibers. The ability to substantially reduce the number of bound modes propagating along the velocity barrier also acts to reduce electronic losses due to scattering, which is enhanced for multimode waveguides supporting many different modes propagating at several different velocities.

\subsection{\label{subssec3}The linear velocity barrier}

\begin{figure}[tb]
 \includegraphics[width=0.4\textwidth]{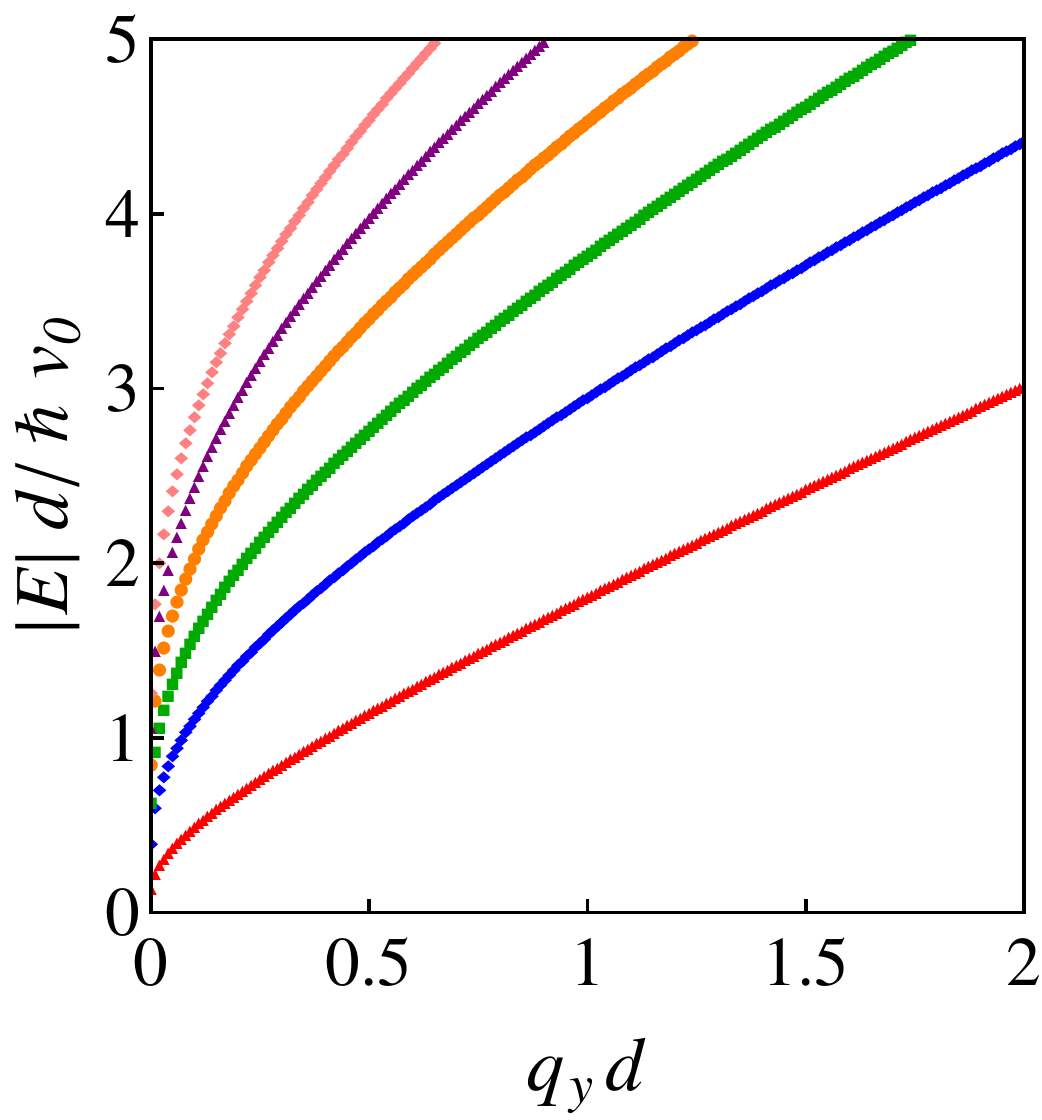}
 \caption{Progression of the six lowest bound states energies with traversal momentum $q_y d$, for massless Dirac particles in a linear velocity barrier, calculated via Eq.~\eqref{eqlinear02}.}
 \label{fig4}
\end{figure}

Now we look at the linear velocity barrier, governed by the parameters $v_0$ and $d$ and shaped by
\begin{equation}
\label{eqlinear00}
 v_F(x)= v_0 (1+|x|/d),
\end{equation}
as sketched as the long-dashed blue line in Fig.~\ref{fig1}. We define the useful dimensionless quantity to measure the energies
\begin{equation}
\label{eqlinear000}
 \gamma = \frac{E d}{\hbar v_0}.
\end{equation}
Proceeding in a similar manner to Sec.~\ref{subssec2}, one finds the following spinor wavefunction in region I $(x>0)$ with the aid of the variable $\xi = 2 q_y d(1+\tfrac{x}{d})$
\begin{multline}
\label{eqlinear01}
 \psi_{I}(x)  = \tfrac{c_{I}}{\sqrt{d}} \left(1+\tfrac{x}{d}\right)^{i \gamma} e^{- q_y x} \times \\
\left(
 \begin{array}{c}
 U\left( 1+i \gamma, 1+ 2 i \gamma, \xi \right) 
  \\ \gamma^{-1} U\left( i \gamma, 1+ 2 i \gamma, \xi \right)
 \end{array}
\right),
\end{multline}
where $c_{I}$ is a normalization constant and $U(\alpha, \beta, \xi)$ is the Tricomi function or confluent hypergeometric equation of the second kind \cite{Gradshteyn}. The solution in region II $(x<0)$ is found by interchanging the top and bottom wavefunction components, and making the replacements $x \to -x$ and $c_I \to c_{II} = \pm c_I$. Ensuring a conserved probability current via Eq.~\eqref{intro01} yields the eigenvalue equation
\begin{equation}
\label{eqlinear02}
 U\left( i \gamma, 1+ 2 i \gamma, 2 q_y d \right) = \pm \gamma U\left( 1+i \gamma, 1+ 2 i \gamma, 2 q_y d \right) ,
\end{equation}
where the $\pm$ is associated with the $\pm$ in the definition of $c_{II}$. Although Eq.~\eqref{eqlinear02} must be solved numerically, it is an exact expression and can be solved with any desired accuracy. The result of such computations is shown in Fig.~\ref{fig4} for the six lowest-lying states as a function of transversal wavevector $q_y d$. Most noticeable is the absence of threshold bound state energies as $q_y d \to 0$, markedly different from the sharper, exponential cusp profile encountered in Sec.~\ref{subssec2}. Instead, there is a plethora of bound states with vanishing transversal momenta. This model suggests that the fundamental change from an exponentially to a linear algebraically growing velocity barrier manifests itself in the loss of the finite threshold effect, which may be important for the design of velocity barriers as guides of electron waves. 

\subsection{\label{subssec4}The square root velocity barrier}

\begin{figure}[t]
 \includegraphics[width=0.4\textwidth]{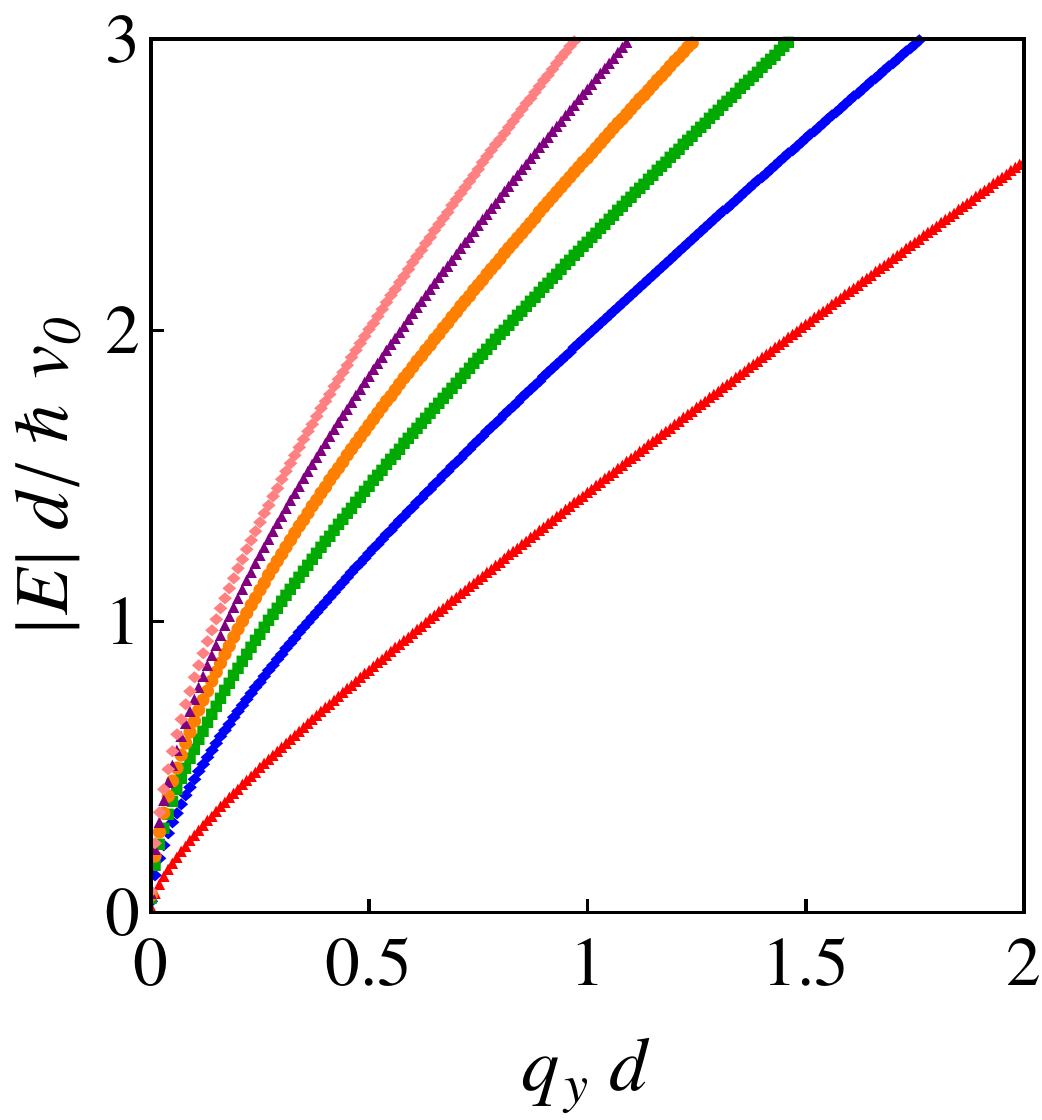}
 \caption{Progression of the six lowest bound states energies with traversal momentum $q_y d$, for massless Dirac particles in a square root velocity barrier, calculated via Eq.~\eqref{eqsqrt02}.}
 \label{fig5}
\end{figure}

Finally, we consider a weakly growing square root velocity barrier, graphed in Fig.~\ref{fig1} as the dot-dashed green line, and with the functional form
\begin{equation}
\label{eqsqrt00}
 v_F(x)= v_0 \sqrt{1+|x|/d}.
\end{equation}
One may obtain the spinor wavefunction in region I $(x>0)$ in an analogous fashion to Sec.~\ref{subssec3}. We obtain, with the assistance of the variable $\xi = 2 q_y d(1+\tfrac{x}{d})$, the solution
\begin{multline}
\label{eqsqrt01}
 \psi_{I}(x)  = \tfrac{c_{I}}{\sqrt{d}} \left(1+\tfrac{x}{d}\right)^{1/2} e^{- q_y x} \times \\
\left(
 \begin{array}{c}
 U\left( 1- \tfrac{\gamma^2}{2 q_y d}, \tfrac{3}{2}, \xi \right) 
  \\ \gamma^{-1} \left( 2 q_y d \right)^{1/2} U\left( \tfrac{1}{2} - \tfrac{\gamma^2}{2 q_y d}, \tfrac{3}{2}, \xi \right)
 \end{array}
\right),
\end{multline}
where $\gamma$ is defined in Eq.~\eqref{eqlinear000} and $c_{I}$ is fixed via normalization. The solution in region II $(x<0)$ is found by interchanging the top and bottom wavefunction components, and making the replacements $x \to -x$ and $c_I \to c_{II} = \pm c_I$. The eigenvalues are wholly governed by the expression
\begin{equation}
\label{eqsqrt02}
 \gamma^{-1} \left( 2qd \right)^{1/2} U\left( \tfrac{1}{2} - \tfrac{\gamma^2}{2 q d}, \tfrac{3}{2}, 2qd \right) = \pm U\left( 1- \tfrac{\gamma^2}{2 q d}, \tfrac{3}{2}, 2qd \right) ,
\end{equation}
which is tractable with standard root-searching procedures. Solutions of Eq.~\eqref{eqsqrt02} are shown in Fig.~\ref{fig5} for the six lowest-lying states. As was the case for the linear barrier in Sec.~\ref{subssec3}, the system observes the vanishing zero-point energy phenomenon as $q_y d \to 0$, which is a hallmark of algebraic velocity barriers growing at most linearly.

Upon comparing the corresponding energy level versus transverse momentum dependences in Fig.~\ref{fig3}, Fig.~\ref{fig4} and Fig.~\ref{fig5} one notices as the velocity barrier growth becomes shallower the bound state energies lower and the inter-energy spacing reduces. As soon as the velocity barrier is growing asymptotically linearly, there is no longer a threshold effect as $q_y d \to 0$ for the zero-point energy, a characteristic which persists for sub-linear velocity barriers. These features are important for the design of few electron mode guiding devices, which require robust on and off logical states.

We should mention that we mostly considered velocity barrier models growing asymptotically. If these models were modified such that the spatial profile of $v_F(x)$ instead saturated at some large but finite value, the corresponding  expressions presented in this work [c.f. Eq.~\eqref{eqcusp02}, Eq.~\eqref{eqlinear02} and Eq.~\eqref{eqsqrt02}] will slightly overestimate the magnitudes of the bound state energies. However, the low-lying states, which are the main focus of this work, are unaffected for all practical purposes by this lack of a saturation.

\section{\label{sec3}Radial velocity barriers}

It is worthwhile to also consider axially symmetric velocity barriers $v_F = v_F(r)$ with the Hamiltonian~\eqref{intro00} to try to achieve total (0D) confinement. We separate the variables in polar coordinates $(r, \theta)$ with the ansatz $\Psi (r, \theta) = (2 \pi)^{-1/2} \left[ e^{i m \theta} \psi_1(r), i e^{i (m+1) \theta} \psi_2(r) \right]^T$. Here $m=0,\pm1, \pm2...$ is related to angular momentum quantum number. Explicitly, the spinor wavefunction satisfies $J_z \Psi = (m+1/2) \Psi$, where the total angular momentum operator $J_z = -i \hbar \partial_{\theta} + \hbar \sigma_z /2$.

Now, a finite circular velocity barrier cannot trap particles (unlike the equivalent 1D case described in Sec.~\ref{subssec1}) because the wavefunctions are always described in terms of standard Bessel functions. These functions are non-square integrable, as was already encountered in the case of massless Dirac fermions in an electrostatic barrier \cite{Chaplik}, since they map onto the scattering states of the Schrodinger equation. However, an infinite sharp velocity barrier can lead to bound states, as we shall see in Sec.~\ref{subssec5}, and more significantly so too does an algebraically smooth velocity barrier model of the inverted Lorentzian type \cite{Downing}, which is shown to be integrable in Sec.~\ref{subssec6}. Both of these aforementioned models exhibit a finite zero-point (ground state) energy. Furthermore, we note there is a general property that states with $m = 0, -1$ are marginally non-square-integrable in the auxiliary spinor $\Psi$ and so may correspond to extended states (rather than bound states) in the full spinor $\Phi$ depending on the asymptotics of the velocity barrier.

\subsection{\label{subssec5}The infinite velocity barrier}

\begin{figure}[tb]
 \includegraphics[width=0.4\textwidth]{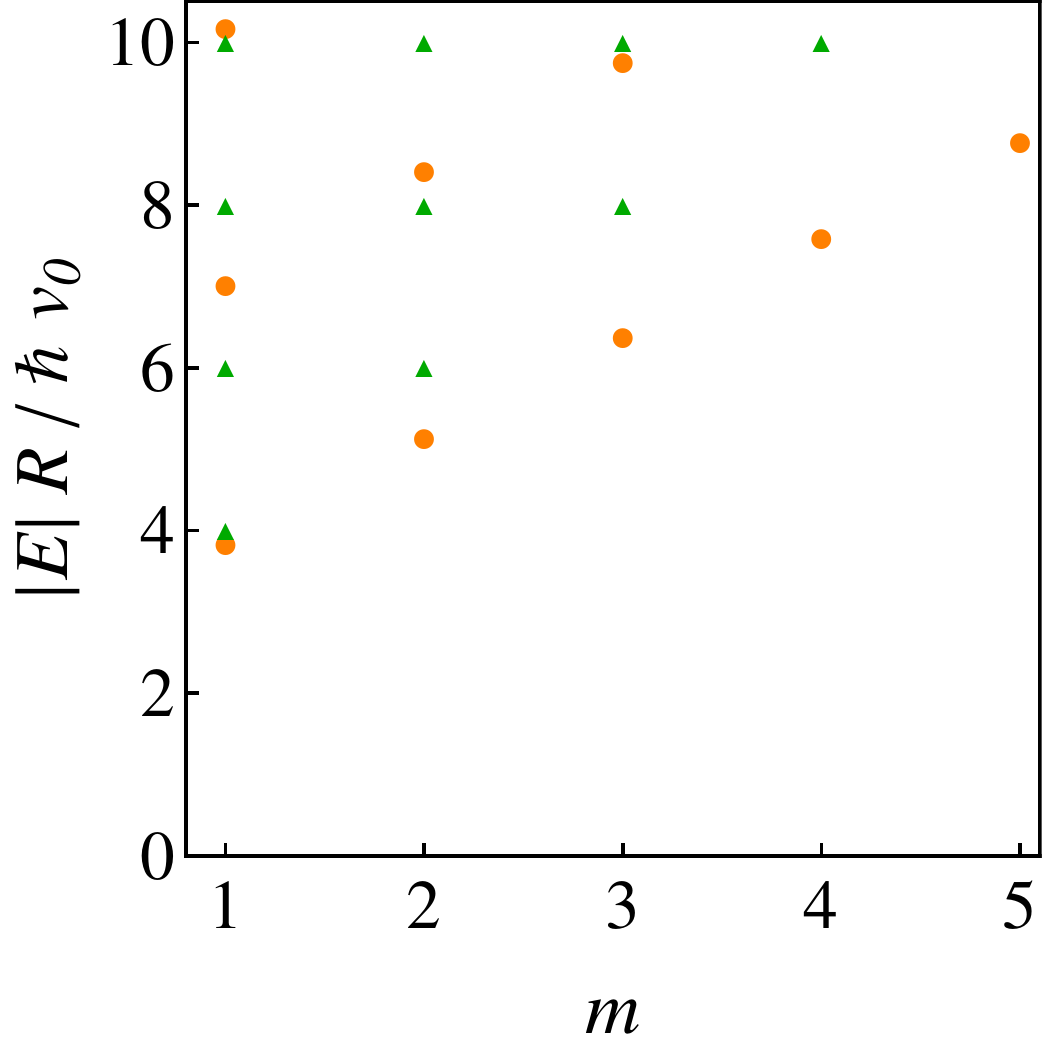}
 \caption{(Color online) Energy levels as a function of $m$ for the infinite velocity barrier (orange circles) calculated via Eq.~\eqref{squarecircle2}; and the inverted Lorentzian velocity barrier (green triangles) calculated via Eq.~\eqref{eqsquare02333333333}.}
 \label{fig6}
\end{figure}

The simplest integrable model is the infinite velocity barrier, defined using the radial distance $R$ as
\begin{equation}
\label{squarecircle1}
 v_F(r)= \begin{cases} v_0, \quad r \le R,  \\
                     \infty, \quad r > R.  \end{cases}
\end{equation}
The Hamiltonian~\eqref{intro00} with this sharp velocity barrier leads to a countably infinite number of bound states, described by 
\begin{equation}
\label{squarecircle2}
 E_{n, m} = \pm \frac{\hbar v_0}{R} \alpha_{\sigma, n}, \quad \sigma = 
  \begin{cases} 
   m, & \text{if } m \ge 1, \\
   -m-1,  & \text{if } m \le -2.
  \end{cases}
\end{equation}
where $\alpha_{\sigma, n}$ is the $n$th positive zero of the Bessel function of the first kind, satisfying $ J_{\sigma}(\alpha_{\sigma, n}) = 0$. It is understood that the eigenvalues respect two symmetries: firstly, for every solution with $E$ there is a solution with $-E$; and secondly the eigenvalues are degenerate with the quantum number replacement $m \to -(m+1)$. The energy levels as a function of angular momentum $m$ are plotted in Fig.~\ref{fig6} as orange circles. How bound states with increasingly high angular momenta appear at higher energies and the threshold for the first bound state to appear, are both features which can be clearly seen. The explicit threshold energy at which the first bound state appears is $E_{1, 1} = E_{1, -2} \simeq 3.83 \hbar v_0/R$.

Notably the $m = 0, -1$ states are associated with non-square integrable auxiliary spinors $\Psi$, which is a common feature for all radial problems of this type (namely those with velocity barriers strengths tending towards infinity as $r \to \infty$). This is because these states have minimal angular momentum and so are highly susceptible to the Klein tunneling phenomenon of perfect transmission at normal incidence. Nevertheless, these extended states are associated with radial wavefunctions which do decay algebraically ($\sim 1/r$), and so may be important for studies of resonant scattering. These special modes appear at the degenerate level $E_{1, 0} = E_{1, -1} \simeq 2.40 \hbar v_0/R$ in the infinite velocity barrier.

\subsection{\label{subssec6}The inverted Lorentzian velocity barrier}

Let us now consider a smooth, algebraic model given by the inverted Lorentzian
\begin{equation}
\label{eqsquare00}
 v_F(r)= v_0 (1+r^2/R^2).
\end{equation}
Working in the variable $\xi = r^2/R^2$, one can construct the radial part of the spinor wavefunction $\psi(r)$ in terms of Gauss hypergeometric functions
\begin{multline}
\label{eqsquare01}
 \psi(r)  = \tfrac{c}{R} \left(1+\xi\right)^{-p_m/2} \times \\
\left(
 \begin{array}{c}
 \xi^{\tfrac{|m|}{2}}  {}_2 F_1 \left( \left[\tfrac{p_m}{2} + \tfrac{E R}{2\hbar v_0} \right], \left[\tfrac{p_m}{2} - \tfrac{E R}{2\hbar v_0} \right], 1+|m|, \tfrac{\xi}{1+\xi} \right) 
  \\ \Upsilon \xi^{\tfrac{|1+m|}{2}}  {}_2 F_1 \left( \left[\tfrac{p_m}{2} + \tfrac{E R}{2\hbar v_0} \right], \left[\tfrac{p_m}{2} - \tfrac{E R}{2\hbar v_0} \right], p_m-|m|, \tfrac{\xi}{1+\xi} \right) 
 \end{array}
\right)
\end{multline}
where $c$ is a normalization constant and where we have made use of the following number
\begin{equation}
\label{eqsquare023434}
 p_m = 1 + |m| + |1+m|,
\end{equation}
and with the prefactor $\Upsilon = \tfrac{1}{2(1+m)} \tfrac{E R}{\hbar v_0}$ when $m\ge0$ and $\Upsilon = 2m \tfrac{\hbar v_0}{E R}$ when $m<0$ respectively. One notices the low angular momentum states $m=0, -1$ are (marginally) non-square integrable in the auxiliary spinors $\Psi$ from the $r \to \infty$ asymptotics of the radial wavefunction, as was foreshadowed in Sec.~\ref{subssec5}. These extended states, which reside at the degenerate level $E_{0, 0} = E_{0, -1} = \pm 2 \hbar v_0/R$, are fully normalizable in the full spinor $\Phi$ due to the presence of the function Eq.~\eqref{eqsquare00}.

Terminating the hypergeometric function in Eq.~\eqref{eqsquare01} by setting either of the first two arguments of the hypergeometric function to be a negative integer or zero, one finds the eigenvalues of the system
\begin{equation}
\label{eqsquare02333333333}
 E_{n, m} = \pm \left( 2n + p_m \right) \frac{\hbar v_0}{R}, \quad n=0,1,2...
\end{equation}
Fig.~\ref{fig6} displays as green triangles the energy levels as a function of $m$. Most notable is the threshold bound state energy of $E_{0, 1} = E_{0, -2} = \pm 4 \hbar v_0/R$ at which the first confined mode with $m \ne 0, -1$ is found. This confirms a finite zero-point energy arises even for this smooth, algebraically growing velocity barrier model and is not an artefact of the minimal model of Sec.~\ref{subssec5}.

These proposals for truly bound states in radial velocity barriers joins a small list of setups which can confine massless Dirac fermions in quantum-dot-like systems. Recent experimental work has focused on combined electric and magnetic fields \cite{Freitag} and electron whispering gallery modes \cite{ZhaoY}, rather than Fermi velocity-induced effects.

\subsection{\label{subssec66}Zero-energy states in radial velocity barriers}

Zero-energy states associated with Dirac Hamiltonians are of great interest due to their importance for topological and Majorana physics. Radial velocity barriers also admit zero-energy state ($E=0$) solutions, which form the degenerate ground state of the system. We consider velocity barriers with the short range behavior $v_F(r \sim 0) \sim r^0$, such that the spinor solutions to the eigenproblem~\eqref{intro00} for nonnegative $m$ take the form
\begin{multline}
\label{eqzero01}
 \Phi  \propto 
\left(
 \begin{array}{c}
 e^{i m \theta} r^m / \sqrt{v_F(r)}
  \\ 0
 \end{array}
\right), \quad m = 0, 1, 2...
\end{multline}
In order to be a normalizable solution for a certain state $m$, the velocity barrier must grow asymptotically faster than $v_F(r \to \infty) \sim r^{2 (1+m)}$, limiting the degeneracy of the ground state. Meanwhile, the eigenvector for negative $m$ is given by
\begin{multline}
\label{eqzero02}
 \Phi  \propto  
\left(
 \begin{array}{c}
 0
  \\ e^{i (m+1) \theta} r^{-(m+1)} / \sqrt{v_F(r)}
 \end{array}
\right), ~ m = -1, -2, -3...
\end{multline}
which is square integrable for a state with quantum number $m$ as long as the velocity barrier grows at large distances more rapidly than $v_F(r \to \infty) \sim r^{-2 m}$. These wavefunctions~\eqref{eqzero01}~and~\eqref{eqzero02} display the chiral property of total suppression of the electronic probability density on one of the the two sublattices, dependent on the sign of the angular momentum $m$. 

\section{\label{conc}Conclusion}

We have studied the appearance and nature of bound states of 2D massless Dirac fermions, which is a nontrivial task due to the phenomena of Klein tunneling, that may arise in several different velocity barrier configurations, including trench-like and radial geometries. We have shown how velocity barrier channels growing linearly or sub-linearly support bound modes for arbitrarily small transversal wavevectors, whereas  algebraically faster (or indeed exponentially) growing barriers posses a finite zero-point energy. This geometry is a candidate to observe the ballistic guiding of few-mode electronic waves. In contrast, a radial velocity barrier growing algebraically has been shown to have a threshold energy at which the first bound state appears, as has a simple circular radial barrier model. Extended states with quantum number $m = 0, -1$ are not always square-integrable but do decay algebraically and so may be consequential for resonant scattering.

These results open up an intriguing avenue to explore in the ongoing quest to achieve trapping and guiding of massless Dirac particles \cite{Lee2016, Gutierrez2016}. With the ongoing improvements in Fermi velocity engineering, particularly via the fabrication of Dirac materials embedded in various substrates \cite{Hwang2012} and controllably strained devices \cite{Downs2016}, we hope that velocity waveguides and traps, as well as the predicted threshold behavior and confinement-deconfinement transitions of the bound states, can be demonstrated in the laboratory in the near future.

\section*{Acknowledgments}
We acknowledge financial support from the CNRS, as well as the EU H2020 RISE project CoExAN (Grant No. H2020-644076), EU FP7 ITN NOTEDEV (Grant No. FP7-607521), and the FP7 IRSES projects CANTOR (Grant No.~FP7-612285), QOCaN (Grant No. FP7-316432), and InterNoM (Grant No. FP7-612624). We would like to thank Jenny Zhao for several illuminating discussions.

\end{document}